\documentclass[useAMS,usenatbib,usegraphicx]{mn2e}
\makeatletter
\def\endfigure{\end@float}
\def\endfigure{\end@float}
\makeatother
\usepackage{afterpage}
\usepackage{float}
\usepackage{url}
\usepackage{subfigure}

\usepackage{amsmath,amssymb}
\usepackage{color}
\usepackage{mathpazo}
\def\vec#1{\mbox{\boldmath $#1$}}
\begin{document}
\title{Measuring the mass distribution of voids with stacked weak lensing }
\author[Y.Higuchi, M.Oguri and T.Hamana]{Yuichi Higuchi$^{1}$\thanks{E-mail: yuichi.higuchi@nao.ac.jp}, Masamune Oguri$^{2}$ and Takashi Hamana$^{3}$\\
$^{1}$Department of Astronomy, Graduate school of Science, University of Tokyo, Tokyo 113-0033, Japan\\
$^{2}$Kavli Institute for the Physics and Mathematics of the Universe (Kavli IPMU, WPI), University of Tokyo, Chiba 277-8583, Japan\\
$^{3}$National Astronomical Observatory of Japan, Mitaka, Tokyo 181-8588, Japan}
\maketitle
\begin{abstract}
We study the prospects for measuring the dark matter distribution of
voids with stacked weak lensing. We select voids from a large set of
$N$-body simulations, and explore their lensing signals with the full
ray-tracing simulations including the effect of the large-scale
structure along the line-of-sight. The lensing signals are compared with
simple void model predictions to 
%TH% reconstruct
infer the three-dimensional mass
distribution of voids. We show that the stacked weak lensing signals are
detected at significant level (S/N$\geq5$) for a $5000$ degree$^2$
survey area, for a wide range of void radii up to $\sim50$ Mpc. The
error from the galaxy shape noise little affects lensing signals at large
scale. It is also found that dense ridges around voids have a great
impact on the weak lensing signals, suggesting that proper modeling of
the void density profile including surrounding ridges is essential for
extracting the average total underdens mass of voids.   
\end{abstract}

% look up list of allowable keywords for this section

%%%%%%%%%%%%%%%%%%%%%%%%%%%%%introduction%%%%%%%%%%%%%%%%%%%%%%%%%%%%%%%%%%
\section{Introduction}
Cosmological structure formation and matter distribution of the universe
depend on models of cosmology. Therefore one of the keys for
constraining cosmology is to understand the distribution of
matter. High density peaks in the matter density distribution which
correspond to galaxies and galaxy clusters have been well studied
\citep{2005ApJ...619L.143B, 2011ApJ...738...41U,
  2012MNRAS.420.3213O}. These overdensity regions are used for
constraining cosmology \citep{2006PhRvD..74d3505T, 2011PhRvD..83b3008O, 2012MNRAS.423.3018P}. 
On the other hand, negative density troughs which correspond to voids have attracted less attention in spite of its long history \citep{1981ApJ...248L..57K} and the fact that they fill more than $80\%$ of the volume of the universe and therefore are an essential component of large scale structure (LSS) of the universe.

There has been analytical work that studied the formation and evolution of voids \citep{1984ApJ...281....9F, 1984PThPh..71..938S, 2006MNRAS.366..467F}. The early evolution of void is well described by a spherical collapse model. In this model underdense regions expand and gradually become deep until the shell crossing occurs \citep{1980lssu.book.....P}. Effects of different cosmology  on voids have also been studied \citep{2006JCAP...05..001M, 2007PhRvL..98h1301P, 2009ApJ...696L..10L, 2009JCAP...01..010K, 2012MNRAS.421.3481L}. These papers pointed out possibilities for constraining cosmology with voids, including primordial non-gaussianity and modified gravity theories.

On the other hand, observationally various void finding algorithms have been applied to redshift surveys such as the Sloan Digital Sky Survey (SDSS) to study properties of voids \citep{1992ApJ...385..421S, 2004MNRAS.355..747J, 2008MNRAS.387..933C, 2012MNRAS.421..926P}. It was found that voids are significantly underdense in their observations which use galaxy distribution for locating voids. These results are broadly consistent with theoretical predictions. 
However many problems still exist when comparing simulations with
observations (e.g. \citealt{1984ApJ...287L..59R, 1999ApJ...522...82K,
  2012arXiv1210.2432T}). 
%TH% In addition, 
For example, the matter distribution of voids
is estimated by assuming the galaxy bias, and therefore the uncertainty
of the bias is one of the most serious challenges in void studies. 

By contrast, gravitational lensing traces all matters regardless of baryon
or dark matter. In fact, the possibility of studying voids with weak
gravitational lensing was discussed in \citet{2000ApJ...530L...1J}, and
a possible statistical detection was reported by
\citet{2002ApJ...580L..97M}. While lensing signals of individual voids
are weak \citep{1999MNRAS.309..465A}, 
%TH%
%next generation wide field surveys
%such as Subaru Hyper Suprime-cam survey (HSC) \citep{2006SPIE.6269E...9M} will
%be able to detect such signals and study the true matter distribution in
%voids without any assumptions on the galaxy bias. 
stacked weak lensing technique may enable to detect such signals and
allows us to study the true matter distribution in voids without any
assumptions on the galaxy bias.
This is exactly what we explore in this paper with paying a special
attention to its prospect with next generation wide field surveys
such as Subaru Hyper Suprime-cam survey (HSC)
\citep{2006SPIE.6269E...9M}.

This paper is organized as follows. In Sec.\ref{sec:basicsofanalysis}, we describe basics of our analyzing techniques, focusing on how we can obtain information of dark matter distribution from gravitational lensing. In Sec.\ref{sec:simulation}, we describe our simulation and void finding algorithm. In Sec.\ref{pressschechter}, we describe the mass function of voids. The void model used for fitting lensing signals is presented in Sec.\ref{sec:voidmodel}. In Sec.\ref{sec:result}, we show the result of the stacking analysis and fitting with our model.  We summarize our results in Sec.\ref{sec:conclusion}.  

%%%%%%%%%%%%%%%%%%%%%%%%%%%%%weak lensing method%%%%%%%%%%%%%%%%%%%%%%%%%%%%%%
\section{Basics of analysis}
\label{sec:basicsofanalysis}
%gravitational lensing%
\subsection{basics of gravitational lensing}
\label{sec:basicsofgravitationallensing}
Throughout this paper, we assume a spatially flat
universe. Gravitational lensing effects on the $\alpha$-$\beta$ plane are
characterized by an isotropic stretching called convergence
$\kappa(\vec{\theta})$ and anisotropic distortion called shear
$\gamma_1(\vec{\theta})$ and $\gamma_2(\vec{\theta})$. They are related
to the two dimensional analogue of the Newtonian gravitational potential
$\psi(\vec{\theta})$ as  
\begin{eqnarray}
\label{potential1}
\hspace{2cm}\displaystyle\kappa(\vec{\theta})&=&\frac{1}{2}\nabla^2\psi(\vec{\theta}),\\
\label{potential2}
\displaystyle\gamma_1(\vec{\theta})&=&\frac{1}{2}\left(\frac{\partial^2\psi(\vec{\theta})}{\partial\theta_\alpha^2}-\frac{\partial^2\psi(\vec{\theta})}{\partial\theta_\beta^2}\right),\\
\displaystyle\gamma_2(\vec{\theta})&=&\frac{\partial^2\psi(\vec{\theta})}{\partial\theta_\alpha\partial\theta_\beta},
\label{potential3}
\end{eqnarray}

\begin{figure}
\begin{center}
\includegraphics[width=7cm, bb= 0 0 842 595]{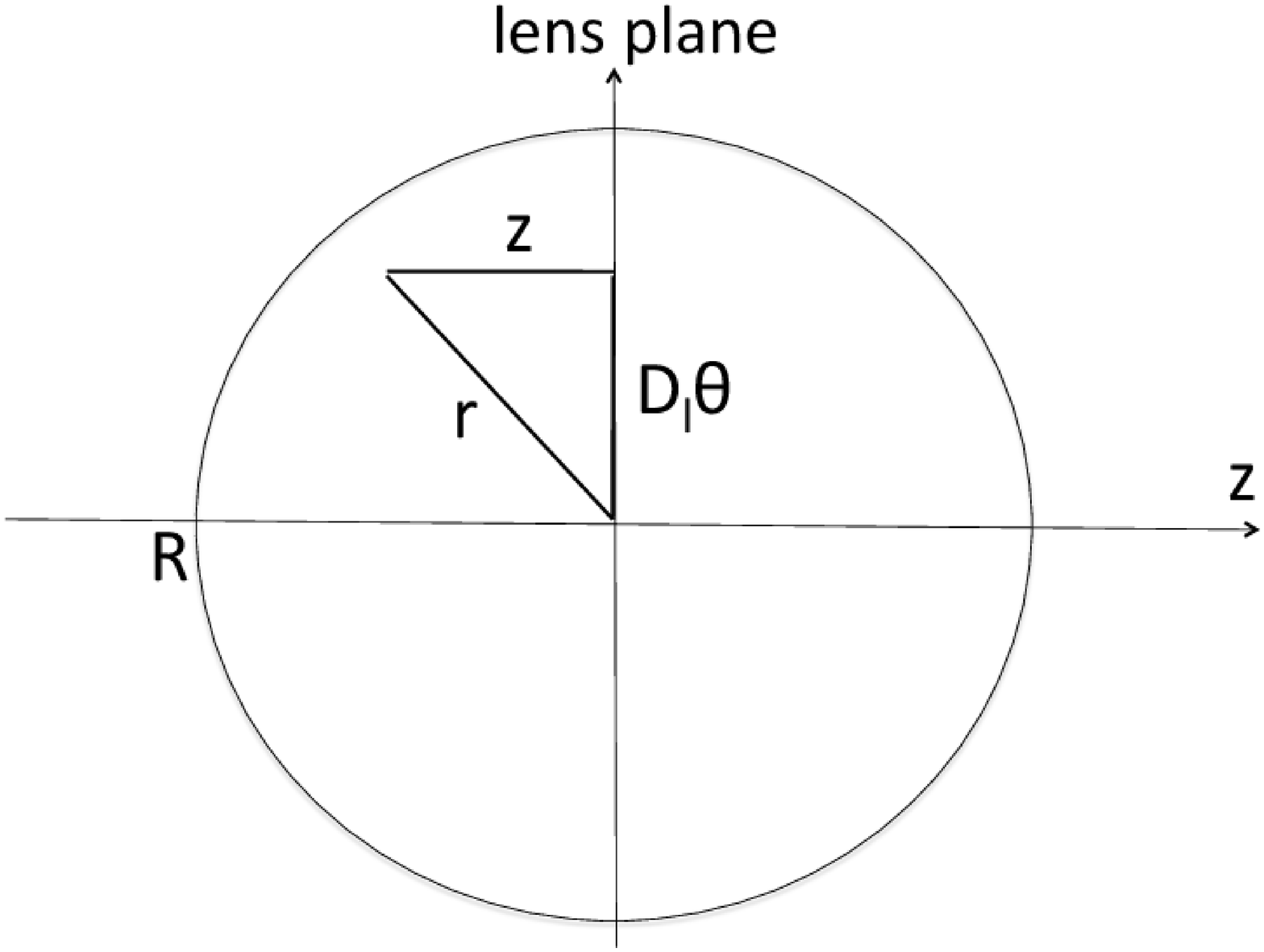}
\caption{Definition of the coordinate system}
\label{coordinate}
\end{center}
\end{figure}

Then convergence at an angular position $\vec{\theta}$ is represented as 
\begin{equation}
\kappa(\vec{\theta})={\frac{3{\mathrm{H_0}}^2\Omega_{\mathrm{m0}}}{2\mathrm{c}^2}}{\frac{\mathrm{D_{ls}D_l}}{\mathrm{D_s {a_l}}^3}}\int d\mathrm{z} \delta(\vec{\theta},\mathrm{z}),
\label{convergence1}
\end{equation}
where z is the line-of-sight distance (fig.\ref{coordinate}), 
% from a lens plane, and $\theta$
%denotes the angular distance from a lens centre;
%\begin{equation*}
%\theta^2=\theta_\alpha^2+\theta_\beta^2.
%\end{equation*}
%Thus the distance r from the lens centre is given by
%(fig.\ref{coordinate}), 
%\begin{equation*}
%\mathrm{r^2=(D_l}\vec{\theta})^2+\mathrm{z}^2.
%\end{equation*}
$\delta(\vec{\theta},z)$ is density contrast defined by
\begin{equation}
\delta(\vec{\theta},z)=\frac{\rho(\vec{\theta},z)-\bar\rho(z)}{\bar\rho(z)},
\end{equation}
with $\bar\rho(z)$ being the average density at z.
$\mathrm{H_0}$ and $\Omega_{\mathrm{m0}}$ are hubble parameter and
matter density parameter at present, respectively, and $\mathrm{a_l}$ is
scale factor at the lens position. $\mathrm{D_s}$, $\mathrm{D_l}$ and
$\mathrm{D_{ls}}$ are angular diameter distance which are expressed as 
\begin{equation}
\mathrm{D_s} =\mathrm{a_s}\chi_\mathrm{s}, \mathrm{D_l =a_l}\chi_\mathrm{l}, \mathrm{D_{ls}} =\mathrm{a_s} (\chi_\mathrm{s}-\chi_\mathrm{l}),
\label{angular}
\end{equation}
where $\mathrm{a_s}$ is the scale factor at a source
position. $\chi_\mathrm{s}, \chi_\mathrm{l}$ are comoving distance from
an observer to source and lens.  
%\afterpage{\clearpage}

It is useful to introduce the tangential distortion component $\gamma_+$ with
respect to a certain point, and the cross component
$\gamma_\times$ which is rotated by $45^\circ$. 
The tangential shear components include all information of lens,  
if a lensing mass profile is axisymmetric on the sky. These two
components are defined with $\gamma_1$ and $\gamma_2$ as  
\begin{equation}
\left(
\begin{array}{cc}
\gamma_+   \\
\gamma_\times     
\end{array}
\right)
=
\left(
\begin{array}{cc}
-\mathrm{cos}2\eta&-\mathrm{sin}2\eta \\
-\mathrm{sin}2\eta& \mathrm{cos}2\eta \\
\end{array}
\right)
\left(
\begin{array}{c}
\gamma_1\\
\gamma_2   
\end{array}
\right),
\label{angle}
\end{equation}
where $\eta$ is the angle between axis-$\alpha$ and $\vec{\theta}$. The
sign of tangential shear 
%TH% indicate that
is chosen so that a background galaxy shape
becomes tangentially deformed for positive and radially deformed for
negative with respect to the lens center.  

 Weak lensing signal is so small that it is difficult to measure masses down to low mass objects. However, by stacking lensing signals from many samples, we can reduce errors which limit to obtain information on the objects (e.g. \citealp{2006MNRAS.372..758M, 2010PASJ...62..811O, 2012MNRAS.420.3213O}). Therefore stacked lensing is a powerful tool for measuring average dark matter distributions of targets. The stacked lensing technique is also used to constrain cosmological parameters (e.g. \citealt{2011PhRvD..83b3008O,2010ApJ...708..645R}). In the weak lensing limit, the average tangential shear profile at $\theta_i$ is related to the convergence \citep{2001PhR...340..291B} as
\begin{equation}
\langle\gamma_+\rangle(\theta_\mathrm{i})=\bar{\kappa}(\theta<\theta_\mathrm{i})-\langle\kappa\rangle(\theta_\mathrm{i}),
\label{tangentialshear}
\end{equation}
where $\langle\cdots\rangle$ denotes average value in the circular annulus, the first term is the mean convergence within a circular aperture of radius $\theta_\mathrm{i}$ defined as 
\begin{equation}
\bar{\kappa}(\theta<\theta_\mathrm{i})=\frac{1}{\pi\theta_\mathrm{i}^2}\int_{{\theta}\leq\theta_\mathrm{i}}{d{\mbox{\footnotesize\vec{\theta'}}}}\kappa({\mbox{\footnotesize\vec{\theta'}}}),
\end{equation}
 and the second term is the mean convergence in the i-th radial bin.
The average cross shear component
$\langle\gamma_{\times}\rangle(\theta<\theta_i)$ must be zero in weak
lensing
%TH%
caused by the standard scalar gravitational potential.

In stacked lensing method, the following statistical uncertainties on tangential shear should be considered
\begin{equation}
\sigma^2_{\gamma_+}=\sigma^2_{\mathrm{void}}+\sigma^2_{\mathrm{LSS}}+\sigma^2_{\mathrm{shape}},
\end{equation}
where $\sigma^2_{\mathrm{void}}$ is the statistical error coming from
deference in the structure of each void used in stacking analysis,
$\sigma^2_{\mathrm{LSS}}$ is the error coming from large-scale structure
along the line-of-sight, and $\sigma^2_{\mathrm{shape}}$ is the shape
noise error coming from the intrinsic ellipticity of galaxies used for
weak lensing measurements. Considering errors coming from structure of
voids and LSS, the covariance matrix between i-th and j-th radial bin is
estimated as   
\begin{equation}
\sigma_{\gamma_+(\theta_\mathrm{i}, \theta_\mathrm{j})}^2=\frac{\displaystyle\sum_{l}^{\mathrm{N_v}}(\gamma_{+,l}(\theta_\mathrm{i})-\langle\gamma_+(\theta_\mathrm{i})\rangle)(\gamma_{+, l}(\theta_\mathrm{j})-\langle\gamma_+(\theta_\mathrm{j})\rangle)}{{\mathrm{N_v}}},
\label{covariance1}
\end{equation}
where $\gamma_{+, l}(\theta_\mathrm{i})$ is the tangential shear in the l-th void at i-th radial bin and $\mathrm{N_v}$ is the number of voids. 
%From eq.\ref{covariance1}, the error at i-th bin becomes
%\begin{eqnarray}
%\sigma_{\gamma_+(\theta_\mathrm{i})}^2\hspace{-0.35cm}&=&\hspace{-0.35cm}\frac{1}{{\mathrm{N_v}}}\left\{\displaystyle\sum^{\mathrm{N_v}}_{l}(\gamma_{+, l}(\theta_\mathrm{i})-\langle\gamma_+(\theta_\mathrm{i})\rangle)(\gamma_{+, l}(\theta_\mathrm{i})-\langle\gamma_+(\theta_\mathrm{i})\rangle) \nonumber\right.\\
%&&\hspace{-1.25cm}\left.+2\displaystyle\sum^{\mathrm{N_v}}_{l}\sum^{\mathrm{N}}_{j\neq \mathrm{i}}(\gamma_{+, l}(\theta_\mathrm{i})-\langle%\gamma_+(\theta_\mathrm{i})\rangle)(\gamma_{+, l}(\theta_j)-\langle\gamma_+(\theta_j)\rangle)\right\}
%\label{stack3}
%\end{eqnarray} 
%where $\mathrm{N}$ is number of radial bin.
 In addition to this, the error from intrinsic ellipticity shown in \citet{2000A&A...358...30V} is determined as
\begin{equation}
\sigma_{\mathrm{shape}}^2=\frac{\sigma_\epsilon^2}{\mathrm{N_vn_gS}},
\label{shapenoise}
\end{equation}
where $\sigma_\epsilon$ is the rms amplitude of the intrinsic
ellipticity distribution, $\mathrm{n_g}$ is the number density of
galaxies and S is the area of a bin. Throughout this paper we adopt
$\sigma_\epsilon=0.4$ and $n_g=30$ arcmin$^{-2}$.
%TH% which resemble survey parameters in the HSC survey. 
As we shall show in below, shape noise becomes the dominant error at
only small scale, on the other hand the other errors becomes dominant at
large scale (sec.\ref{signaltonoiselatio}). We ignore other
observational systematics such as imperfections in the telescope and
electric devices of detectors. 

%SN
\subsection{Signal to noise ratio of stacked lensing}
\label{signal to noise ratio of stacked lensing}
A useful way to quantify the observability is to calculate the total
Signal-to-Noise (S/N) ratio over the considered scale. 
%THT Our calculation follows
We follow the definition by \citet{2011PhRvD..83b3008O}. The total S/N is expressed as 
\begin{equation}
\label{totalsn}
\left(\frac{\mathrm{S}}{\mathrm{N}}\right)^2=\sum_{\mathrm{i,j}}\gamma_{+}(\theta_\mathrm{i})\left[{\rm cov}(\gamma_{+}(\theta_\mathrm{i}), \gamma_{+}(\theta_\mathrm{j}))\right]^{-1}\gamma_{+}(\theta_\mathrm{j}),
\end{equation}
where the summation indices i, j run over radial bins. 
%TH% For calculation
%with the tangential shear we estimate S/N for both situations that
%covariance matrix include and does not include the shape noise. For
%calculation with convergence we can replace $\gamma_{+}$ with $\kappa$
%in eq.\ref{totalsn}. The covariance matrix does not include the shape
%noise for this case. 
We evaluate the S/N with covariance matrix with and without the shape
noise component to see the effect of the shape noise.
The S/N is related to the accuracy of the mass estimate. More specifically, they are related as \citep{2011PhRvD..83b3008O}
\begin{equation}
\frac{\Delta \mathrm{M}}{\mathrm{M}}\sim\left(\frac{\mathrm{S}}{\mathrm{N}}\right)^{-1}\left(\frac{d\mathrm{ln}\gamma_+}{d\mathrm{ln}\mathrm{M}}\right)^{-1}.
\label{accuracy}
\end{equation}
 Therefore S/N is proportional to the accuracy for constraining the mass of the object. 
 
%mass reconstruction
\subsection{Mass reconstruction}
\label{massreconstruction}
It is sometimes useful to derive a model independent information  of the mass from weak lensing data (e.g. \citealt{2010PASJ...62..811O, 1996ApJ...469...73S} and \citealt{1997ApJ...482..648S}). Convergence in each circular annulus of radius $\theta$ is related to the projected mass density
\begin{equation}
\kappa(\theta)=\frac{\Sigma(\theta)}{\Sigma_{cr}},
\label{massreconstruct3}
\end{equation}
where  $\Sigma_{\mathrm{cr}}$ is critical projected mass density defined as 
\begin{equation}
\Sigma_{\mathrm{cr}}=\frac{\mathrm{c}^2}{4\pi \mathrm{G}}\frac{\mathrm{D_s}}{\mathrm{D_l} \mathrm{D_{ls}}}.
\end{equation}
The mass is related to convergence \citep{2001PhR...340..291B} as
\begin{equation}
\mathrm{M}(<\theta)=\Sigma_{\mathrm{cr}}\mathrm{D_l}^2\int d\theta'^2\kappa(\theta')\mathrm{U}(\theta'-\theta),
%\mathrm{M}(<\theta)=\sigma_{cr}\int d\theta'^2\kappa(\theta')\mathrm{U}(\theta'-\theta),
\label{massreconstruct4}
\end{equation}
where $\mathrm{U}(\theta)$ is weight function which we adopt $\mathrm{U}(\theta)=1$ in our calculation. When we estimate void masses which are removed from underdensity regions, we calculate those only for the region with $\kappa(\theta)<0$. For the reconstruction from tangential shear, more careful calculation called $\zeta$-statistics \citep{1994ApJ...437...56F, 2000ApJ...539..540C} is needed.  

Furthermore the mass is related to the average mass density $\bar{\rho}$ and void radius R
\begin{eqnarray}
\label{massreconstruct2}
\hspace{2cm}\mathrm{M}&=&\mid\frac{4}{3}\pi \mathrm{R}^3\bar{\rho}-\frac{4}{3}\pi \mathrm{R}^3\rho\mid\\ \nonumber
&=&\mid\frac{4}{3}\pi \mathrm{R}^3(\bar{\rho}-\rho)\mid\\ \nonumber
&=&\frac{4}{3}\pi \mathrm{R}^3\bar{\rho}\mid\delta\mid,
\end{eqnarray}
where we assume the mass density $\rho$ is constant in the region.

%%%%%%%%%%%%%%%%%%%%%%%%%%%%%%%%raytracing simulation%%%%%%%%%%%%%%%%%%%%%%
\section{simulation}
\label{sec:simulation}
\subsection{Ray-tracing simulation}
\label{sec:raytracingsimulation}

We use a large set of ray-tracing simulation which are carried out in
\citet{2009ApJ...701..945S}. 
To carry out $N$-body simulation, they use
the parallel Tree-Particle Mesh code Gadget-2
(\citealp{2005MNRAS.364.1105S} and see  \citealt{2009ApJ...701..945S}
for simulation details). 
The simulation employ $256^3$ particles in a
cubic box of $240h^{-1}$ Mpc on a side. Cosmological parameters are
assumed as hubble parameter $\mathrm{H_0}=73.2$ km/s/Mpc, matter density
$\Omega_\mathrm{m}=0.238$, baryon density $\Omega_\mathrm{b}=0.042$,
dark energy density $\Omega_{\Lambda}=0.762$, equation of state
parameter $\mathrm{w}=-1$, spectral index $\mathrm{n_s}=0.958$ and
variance of density fluctuation with $8$ $h^{-1}$ Mpc
$\sigma_8=0.76$. These cosmological parameters follows WMAP 3-year
results (\citealp{2007ApJS..170..377S}). 
 By using results of $N$-body simulation, they carry out ray-tracing
 simulation with the algorithm described in 
%TH% \citet{2000ApJ...537....1W} and 
\citet{2001MNRAS.327..169H}. 
We refer the reader to the above reference for more details.
%TH% They set source redshift $\mathrm{z_s}$ at $\mathrm{z_s}\simeq1.0$
%and make 200 realizations. 
We use 200 realizations with a fixed source redshift of $\mathrm{z_s}\simeq1.0$
The area of each realization is $5\times5$ degree$^2$. Therefore
 the effective total area is $5000$ degree$^2$. 

%TH% Here we outline the ray-tracing simulation briefly, see \citet{2009ApJ...701..945S} for  more details. To obtain lensing signal, distance between observer and sources is divided into several regions to create 19 lens planes separated by $120h^{-1}$ Mpc in comoving distance. Then the three dimensional density fluctuation in each each region is projected onto each lens plane. Next they trace 2048$^2$ rays backwards from the observer to the source plane. The initial ray directions are set on 2048$^2$ grids with a grid space of $0.15$ arcmin. Ray-tracing realizations of underlying density field are made by randomly shifting the simulation boxes assuming periodic boundary conditions. The lens planes from the same box are shifted in the same way in order to maintain the clustering pattern of mass distribution in one box.
 
%void finder%
\subsection{Void Finder}
\label{sec:voidfinder}

To search for voids in the simulation data, we employ
  the public code Void Finder (\citealp{2009ApJ...699.1252F}), which is
  based on the void finding algorithm developed by 
\citet{2004ApJ...607..751H} and
  \citet{2005ApJ...620..618H}.
%TH%; interested readers are referred to these papers for details. 
We refer the reader to the above references for full details of the
algorithm and its implementation. Here we describe only points which are
specific to this study.

In this study, we use dark matter haloes for tracers of underlying dark
matter distribution.
The dark matter halos are identified in the $N$-body simulations using
friend-of-friend (FOF) algorithm with the linking length of $b=0.2$
(Oguri \& Hamana 2012).
The Void Finder is applied to haloes with masses larger than
$2.2\times10^{12} M_\odot$.
The Void Finder includes some parameters for which we adopt the same
parameters recommended in \citet{2009ApJ...699.1252F}, except for the voids
minimum radius parameter $\xi$. 
We take two values for the minimum radius, $\xi= 5$ and 10 Mpc, to see
the effect of its choice on void finding.
For stacked lensing analysis, we use voids with the radii (which is
defined in the Void Finder) between
$\mathrm{R}=15$ and 45 Mpc and in the redshift range
$0.4 \leq \mathrm{z} \leq 0.6$ because lensing efficiency is expected to be
high in the redshift range. The number of halos within the redshift
range is $2.5\times10^6$. For comparison, we also use voids with 
$15 \leq \mathrm{R} \leq 40$ Mpc in the redshift range of 
$0.1 \leq \mathrm{z} \leq 0.3$. The number of halos used in this redshift
range is $6.0\times10^5$.

 %press schechter theory
 \section{Press-Schechter theory}
 \label{pressschechter}
 
 We use the modified Press-Schechter (PS) formalism (see
 e.g. \citealt{2009JCAP...01..010K}) for predicting number counts of
 voids analytically. 
  
 In the standard PS theory \citep{1974ApJ...187..425P}, regions which
 exceed the spherical collapse linear overdensity
 $\delta_\mathrm{c}=1.689$ 
%TH%become 
are considered to end up a halo. 
%TH%The probability which has the linear density fluctuation $\delta$
%above $\delta_\mathrm{c}$ becomes as 
The probability that a given point lies in a region with the linear
density fluctuation $\delta > \delta_\mathrm{c}$ is 
\begin{equation}
\mathrm{P}\left(\delta>\delta_\mathrm{c}|\mathrm{M}\right)=\frac{1}{\sqrt{\mathrm{2}\pi}}\frac{\sigma(\mathrm{R})}{\delta_\mathrm{c}}\mathrm{exp}\left(-\frac{\delta^2_\mathrm{c}}{2\sigma^2(\mathrm{R})}\right),
\label{ps1}
\end{equation}
where M is a halo mass defined with a radius R as,
\begin{equation}
\mathrm{M}=\frac{4\pi}{3}\bar{\rho}\mathrm{R}^3,
\label{ps2}
\end{equation} 
and $\sigma^2(\mathrm{R})$ is the rms density fluctuation defined as,
\begin{equation}
\sigma^2(\mathrm{R})=\int\frac{dkk^2}{2\pi^2}\mathrm{W}^2(k\mathrm{R})\mathrm{P}(k),
\label{ps3}
\end{equation}
where $\mathrm{W}(k\mathrm{R})$ is the Fourier transform of the window function and $\mathrm{P}(k)$ is the matter power spectrum. Therefore the mass function n(M) is reduced as 
\begin{eqnarray}
\mathrm{n(M)}d\mathrm{M}\hspace{-0.3cm}&&\nonumber\\
&&\hspace{-1.cm}=2\bar{\rho}\left|\mathrm{P}(\delta>\delta_\mathrm{c}|\mathrm{M})-\mathrm{P}(\delta>\delta_\mathrm{c}|\mathrm{M}+d\mathrm{M})\right| \nonumber\\
&&\hspace{-1.cm}=\sqrt{\frac{2}{\pi}}\frac{\bar{\rho}}{\mathrm{M}^2}\left|\frac{d\mathrm{ln}\sigma(\mathrm{R})}{d\mathrm{lnM}}\right|\frac{\delta_\mathrm{c}}{\sigma(\mathrm{R})}\mathrm{exp}\left(-\frac{\delta^2_\mathrm{c}}{2\sigma^2(\mathrm{R})}\right)d\mathrm{M}.%\nonumber\\
\label{ps4}
\end{eqnarray}
 
 In order to estimate the number count of voids based on the modified PS theory,  we replace $\delta_\mathrm{c}$ in eq.\ref{ps4} to linear underdensity $|\delta_\mathrm{v}|$. 
 
 In our calculation, gaussian window function and the CDM transfer
 function by BBKS \citep{1986ApJ...304...15B} are used. 
%TH%Void masses $\mathrm{M_v}$ are estimated from the weak lensing
%signals with eq.\ref{massreconstruct2}. 
For estimating the number count at a given redshift, the rms linear density
fluctuation $\sigma(\mathrm{R})$ is evolved by using the linear growth factor. 
%%%%%%%%%%%%%%%%%%%%%%%%%%%%void model%%%%%%%%%%%%%%%%%%%%%%%%%%%%%%%%%
\section{Void model}
\label{sec:voidmodel}
%Weak lensing signals are computed to two models. One is our model named double top-hat model, the other is a Krause model studied in \citet{2012arXiv1210.2446K}.  

%\subsection{Double top-hat model}
%\label{Doubletophatmodel}
Density profiles of voids have been studied in previous works (e.g. \citealt{2004MNRAS.350..517S, 2012MNRAS.421..926P}), which indicate that the matter density is almost constant over the underdense region and there is a very sharp spike called a ridge at the edge of voids. 

\begin{figure}
\begin{center}
\includegraphics[width=9cm,bb=0 0 842 595]{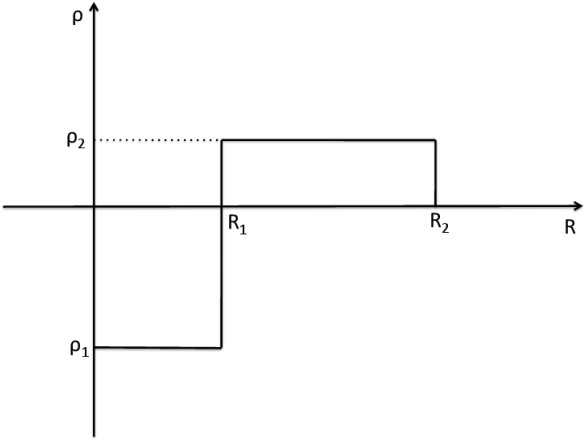}
\caption{Double top-hat void model. In $\mathrm{r<R_1}$,
  $\rho(\mathrm{r})$ is equal to constant $\rho_1$, and
  $\mathrm{R_1<R<R_2}$, $\rho(\mathrm{r})$ is equal to constant
  $\rho_2$. Relative amplitudes between $\rho_1$ and $\rho_2$ are
  determined by the requirement of the mass compensation. $\rho$ is zero
  outside  $\mathrm{R_2}$.  }
\label{doubletophat}
\end{center}
\end{figure}

In order to make a simple void model which includes the properties found in the previous work, we consider a spherically symmetric void model including ridges, which we call a double top-hat model (fig.\ref{doubletophat}).
% In realistic case, there is a structure called ridges which have high mass density outside voids. To reconstruct voids including such structure, we put two top-hats. 
In this model, we set mass density $\rho(r)$ as  
\begin{equation}
\rho(\mathrm{r})= \left\{ 
\begin{array}{ll}
\vspace{0.2cm}\rho_1 & (\mathrm{r\leq R_1}),\\
\vspace{0.2cm}\rho_2 & (\mathrm{R_1<r\leq R_2}),\\
0 & (\mathrm{R_2<r}),
\end{array}
\right.
\label{massdensity1}
\end{equation}
where r is the distance from a void center (fig.\ref{coordinate}), 
$\rho_1$ and $\rho_2$ are constant in each region.
We assume that the total mass between the void region and the ridge
region should 
%TH%follow the mass conservation
be compensated each other. Thus the masses between two regions are related as  
\begin{equation}
\frac{4}{3}\pi{\mathrm{R_1}}^3(\bar{\rho}-\rho_1)=\frac{4}{3}\pi(\mathrm{R_2}^3-\mathrm{R_1}^3)(\rho_2-\bar{\rho}).
\label{masscons}
\end{equation}
From eq.\ref{masscons}, $\rho_2$ is expressed with $\rho_1$, $\mathrm{R_1}$ and $\mathrm{R_2}$
\begin{equation}
\rho_2=\bar{\rho}+\frac{1}{1-(\mathrm{R_2/R_1})^3}(\rho_1-\bar{\rho}).
\label{massdensity2}
\end{equation}
From eq.\ref{massdensity1} and eq.\ref{massdensity2}, density contrast is reduced as 
\begin{equation}
\delta(\mathrm{r})= \left\{ 
\begin{array}{ll}
\vspace{0.2cm}\delta & (\mathrm{r<R_1}),\\
\vspace{0.2cm}\frac{\delta}{1-(\mathrm{R_2/R_1})^3} & (\mathrm{R_1<r<R_2}),\\
0 & (\mathrm{r>R_2}).
\end{array}
\right.
\label{delta2}
\end{equation}
An advantage of this double top-hat model is that its lensing properties can be computed analytically. By using eq.\ref{convergence1} and eq.\ref{delta2}, convergence is calculated as
\begin{eqnarray}
	\kappa(\theta)\hspace{-0.35cm}&=&\hspace{-0.35cm}{\frac{3{\mathrm{H_0}}^2\Omega_{\mathrm{m}0}}{2\mathrm{c}^2}}{\frac{\mathrm{D_{ls}D_l}}{\mathrm{D_s {a_l}^3}}}\int^{\mathrm{D_l}\sqrt{{\theta_2}^2-\theta^2}}_{-\mathrm{D_l}\sqrt{{\theta_2}^2-\theta^2}} d\mathrm{z}\left\{\delta\Theta(\mathrm{R_1-r})\right.\nonumber\\
	&&\left.+\frac{\delta}{1-(\mathrm{R_2/R_3})^3}\Theta(\mathrm{r-R_1})\Theta(\mathrm{R_2-r})\right\} \nonumber\\
%\begin{eqnarray*}	
%	                         &=&{\frac{3{H_0}^2\Omega_{m0}}{2c^2}}{\frac{D_{ls}D_l}{D_s {a_l}^3}}\cdot\left[\left\{\int^{-D_l\sqrt{{\theta_1}^2-\theta^2}}_{-D_l\sqrt{{\theta_2}^2-\theta^2}}dz\frac{\delta}{1-(R_2/R_1)^3}+\int^{D_l\sqrt{{\theta_1}^2-\theta^2}}_{-D_l\sqrt{{\theta_1}^2-\theta^2}}dz\delta\right.\right.\nonumber\\
%	                         &&\hspace{5cm}\left.+\int^{D_l\sqrt{{\theta_1}^2-\theta^2}}_{D_l\sqrt{{\theta_2}^2-\theta^2}}dz\frac{\delta}{1-(R_2/R_1)^3}\right\}\Theta(\theta_1-\theta)\nonumber\\
%	                         &&\hspace{4cm}\left.+\left\{\int^{D_l\sqrt{{\theta_2}^2-\theta^2}}_{-D_l\sqrt{{\theta_2}^2-\theta^2}}dz\frac{\delta}{1-(R_2/R_1)^3}\right\}\Theta(\theta-\theta_1)\Theta(\theta_2-\theta)\right]\nonumber\\
%\end{eqnarray*}	                              
	                         &=&{\frac{3{\mathrm{H_0}}^2\Omega_{\mathrm{m}0}}{2c^2}}{\frac{\mathrm{D_{ls}D_l}}{\mathrm{D_s {a_l}^3}}}\cdot2\delta \mathrm{D_l}\cdot \nonumber\\ 
	                         &&\left[\left\{\frac{\sqrt{{\theta_2}^2-\theta^2}-\sqrt{\theta_1^2-\theta^2}}{1-(\theta_2/\theta_1)^3}+\sqrt{\theta_1^2-\theta^2}\right\}\Theta(\theta_1-\theta)\right.\nonumber\\
	                         &&+\left.\left\{\frac{\sqrt{\theta_2^2-\theta^2}}{1-(\theta_2/\theta_1)^3}\right\} \Theta(\theta-\theta_1)\Theta(\theta_2-\theta)\right],
\label{convergence3}	                         
\end{eqnarray}
where $\Theta(r)$ is a step function. The radii are related to angles 
\begin{eqnarray}
\hspace{3cm}r&=&\mathrm{D_l}\theta,\\
\mathrm{R_i}&=&\mathrm{D_l}\theta_\mathrm{i},
\end{eqnarray}
where the induce i takes 1 or 2. 

In addition to the analytical expression of convergence, we can also calculate the tangential shear analytically. From eq.\ref{tangentialshear}, the tangential shear reduces to
\begin{eqnarray}
\langle\gamma_+\rangle(\theta)&=&\bar{\kappa}(<\theta)-\langle\kappa\rangle(\theta) \nonumber\\
&=&\frac{2}{\theta^2}\displaystyle\int_0^\theta d\theta'\theta'\kappa(\theta')-\langle\kappa\rangle(\theta).
\end{eqnarray}
From eq.\ref{convergence3}, the integration becomes
\begin{eqnarray}
\displaystyle\int^\theta_0d\theta'\theta'\kappa(\theta')&=&\nonumber\\
&&\hspace{-2cm}\left\{\frac{\mathrm{A}}{3}\frac{(\theta_2/\theta_1)^3(\theta_1^2-\theta^2)^{\frac{3}{2}}-(\theta_2^2-\theta^2)^\frac{3}{2}}{1-(\theta_2/\theta_1)^3}\right\}\Theta(\theta_1-\theta)\nonumber\\
&&\hspace{-2cm}-\left\{\frac{\mathrm{A}}{3}\frac{(\theta_2^2-\theta^2)^{\frac{3}{2}}}{1-(\theta_2/\theta_1)^3}\right\}\Theta(\theta-\theta_1)\Theta(\theta_2-\theta),
\label{derivation2}
\end{eqnarray}
where $\mathrm{A}={\frac{3{\mathrm{H_0}}^2\Omega_{\mathrm{m}0}}{\mathrm{c}^2}}{\frac{\mathrm{D_{ls}D_l^2}}{\mathrm{D_s {a_l}^3}}}\delta$. Therefore with eq.\ref{convergence3} and eq.\ref{derivation2}, an analytical expression of the tangential shear is derived as
\begin{eqnarray}
\langle\gamma_+\rangle(\theta)&=&\nonumber\\ 
&&\hspace{-1.5cm}\left[\frac{\mathrm{A}}{3\theta^2}\frac{1}{1-(\theta_2/\theta_1)^3}\left\{(\theta_2/\theta_1)^3(2\theta_1^2+\theta^2)\sqrt{\theta_1^2-\theta^2}\right.\right.\nonumber\\
&&\hspace{-1.5cm}\left.\left.-(2\theta_2^2+\theta^2)\sqrt{\theta_2^2-\theta^2}\right\}\right]\Theta(\theta_1-\theta)\nonumber\\
&&\hspace{-1.5cm}-\left\{\frac{\mathrm{A}}{3\theta^2}\frac{(2\theta_2^2+\theta^2)\sqrt{\theta_2^2-\theta^2}}{1-(\theta_2/\theta_1)^3}\right\}\Theta(\theta-\theta_1)\Theta(\theta_2-\theta).
\label{tangentialshear3}
\end{eqnarray}
We assume the density contrast is equal to zero outside the second top-hat. Therefore the convergence and the tangential shear must be zero in that region.

%TH%We adopt the $\chi^2$ method to extract parameters from simulations. 
Model parameters are determined by the standard $\chi^2$ minimization
method with simulation results.
We define the estimator $\chi^2$ as 
%\begin{equation}
%\chi^2=\sum_i\frac{\left[\gamma_{+, i, \mathrm{model}}(\delta,\theta_1,\theta_2)-\gamma_{+, i, \mathrm{sim}}\right]^2}{\sigma_i^2}
%\end{equation}
\begin{eqnarray}
\chi^2&=&\sum_{i,j}\left(\gamma_{+, \mathrm{model}}(\theta_i)-\gamma_{+, \mathrm{sim}}(\theta_i)\right)\nonumber \\
&&\hspace{0.5cm}\times\left[{\rm cov}(\gamma_{+, \mathrm{sim}}(\theta_i), \gamma_{+, \mathrm{sim}}(\theta_j))\right]^{-1}\nonumber \\
&&\hspace{0.5cm}\times\left(\gamma_{+, \mathrm{model}}(\theta_j)-\gamma_{+, \mathrm{sim}}(\theta_j)\right),
\end{eqnarray}
where $\gamma_{+, \mathrm{model}}(\theta_i)$ and $\gamma_{+, \mathrm{sim}}(\theta_i)$ are tangential shear in the model and simulation at i-th radial bin. 

%In the calculation for fitting simulated lensing signals from voids which exits in the redshift range $z=0.4\sim0.6$, we fix the lens redshift at $\mathrm{z_l}=0.5$. 
%and $\sigma_i$ is 1-sigma error estimated from simulation at i-th radial bin.  

\section{Result}
\label{sec:result}
%numbercount
\subsection{Number counts of voids}
\label{sec:numbercount}
\begin{figure}
\begin{center}
\includegraphics{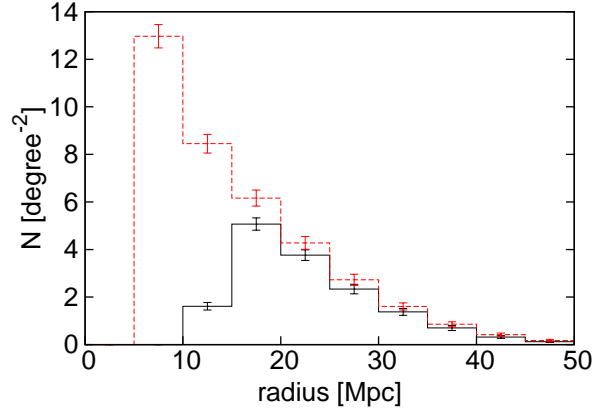}
\caption{Number counts of voids in the simulation for the redshift range $0\leq z\leq 1$. x-axis is void radius. y-axis is number counts of void in each bin. The dashed and solid line histograms show the number counts with minimum radius $\xi=5$ Mpc and $\xi=10$ Mpc. Error bar shows 1$\sigma$ with 200 realizations.}
\label{numbercount1}
\end{center}
\end{figure}

In order to check effects of the minimum radius parameter $\xi$
in the Void Finder on void finding, we run the Void Finder with
both $\xi=5$ Mpc and $\xi=10$ Mpc. 
We compute the void number counts as a function of void radius for the two
cases, and present the results in Figure.\ref{numbercount1}, where voids in the
redshift range $0\leq z \leq1$ are considered.
The figure clearly shows that the number counts depend on the choice of
the minimum radius. 
It is found that the number counts of voids for the lager
minimum radius case is smaller than the other case with two features;
one is the cutoff at smaller scales and the other is the effect originated from the void finding algorithm.
%systematically smaller counts at larger scales.
The reason of the former is a natural consequence of the large minimum
radius of $\xi=10$ Mpc, which prevents to identify voids with the size smaller than that.
%The latter is originated from the algorithm of the Void
%Finder: It first locates small original void and merges connected original voids, then finally defines separate voids. 
%For a larger minimum radius case, less small original voids are identified,
%resulting in a smaller number of final voids.
For the latter, the same trend is found in \citet{2009ApJ...699.1252F}.
As is found in Fig \ref{numbercount1}, these effects are not
significant for larger radii R $\geq20$ Mpc.
It should be also noted that not all the voids are associated with
{\it legitimate} cosmological voids but some are {\it spurious} voids originated from a
lack of halos for those regions to which the Void Finder is adopted. 
The spurious voids are likely mis-identified when a void radius is
similar to or smaller than the mean separation of halos.
In our case, the mean separation of halos is about 7 Mpc, thus voids
with a radius $\geq15$ Mpc might not be seriously affected by this effect regardless of the choice of the minimum radius.
Taking the above points into account, in the following stacking
analysis, we take the minimum radius $\xi=5$ Mpc, and consider voids
with the radius $\geq15$ Mpc. 

%\begin{figure}
%\begin{center}
%\includegraphics{numbercount1.eps}
%\caption{Number counts of voids in simulation and the modified PS theory. X-axis is void radius. Y-axis is number count of void in each bin. Error bar shows 1$\sigma$ with 200 realizations. solid line: the number count in simulation when the minimum radius is set to $\xi=5$ [Mpc], dash line: number count from the PS theory with $\delta_\mathrm{v}=-0.15$, }
%\label{numbercount1}
%\end{center}
%\end{figure}
%\afterpage{\clearpage}

%removed mass with convergence map
\subsection{Direct estimation of void masses}
\label{sec:removedmass}
 In order to estimate mass profiles and void masses, we stack lensing
 signals from voids selected by the Void Finder. We estimate the mass in
 each void sample from the result of stacking and
 eq.\ref{massreconstruct4}.  

%TH% In sec.\ref{sec:numbercount} we obtained the number counts of voids as
% a function of void radius. 
We divide voids located by Void Finder into 6 groups based on
 the void radius determined by the Void Finder. Here we select only voids that
 are located in the redshift range  $0.4 \leq \mathrm{z} \leq0.6$. We align
 the centers of voids determined by the Void Finder and stack those in
 each group. The radii and number of voids used for stacking analysis are
 showed in table.\ref{stackref1}. Points with errors in
 figure.\ref{convergenceprofile1} and
 figure.\ref{tangentialshearprofile2} show stacked convergence and
 tangential shear profiles in each radius.  
 
 The void masses we estimate from the convergence profiles with eq.\ref{massreconstruct4} are presented in table.\ref{stackref1}. The
 calculation is carried out only for the region $\kappa(\theta)<0$. We
 find that void masses that are directly estimated from the convergence
 profiles are comparable to typical cluster masses
 ($\sim10^{14}$M$_\odot$). As we will see below, these void masses are
 smaller than 
%TH%true removed masses of the voids 
ones estimated from the double top-hat model because of the effect of
 surrounding ridges. 
\begin{table}
\begin{center}
\caption{Details of stacking. Column (1): radius of voids determined by
  the void finder; Column(2): the number count of voids in the each
  radial bin and the redshift range $\mathrm{z_l}=0.4\sim0.6$;
  Column(3): void mass calculated by integrating the $\kappa(\theta)<0$
  region. }
\begin{tabular}{ccc}
\hline
Radius [Mpc]&Number of voids&Mass from $\kappa(\theta)<0$ [$\mathrm{M_{\odot}}$]\\ \hline\hline
$15\sim20$ &5246&$9.03\times10^{13}$\\ %\hline
$20\sim25$ &3892&$4.45\times10^{14}$\\ %\hline
$25\sim30$ &2446&$1.81\times10^{15}$\\ %\hline
$30\sim35$ &1400&$3.06\times10^{15}$\\ %\hline
$35\sim40$ &724&$4.01\times10^{15}$ \\ %\hline
$40\sim45$ &320&$3.26\times10^{15}$\\ 
\hline
\label{stackref1}
\end{tabular}
\end{center}
\end{table}
%\begin{figure}
%\begin{center}
%\includegraphics[width=10cm, bb=0 0 720 504]{mstackedprofile_6.pdf}
%\caption{convergence profile of void and LSS in simulation. red: $15<R<20$ [Mpc]. green: $20<R<25$ [Mpc]. blue: $25<R<30$ [Mpc]. pink: LSS. %Vertical line shows void radius which are decided in void finder. Error bar shows 1$\sigma$.}
%\label{stackkap}
%\end{center}
%\end{figure}
%\begin{figure}
%\begin{center}
%\includegraphics[width=10cm, bb=0 0 720 504]{stack_conv.pdf}
%\caption{convergence profile of void in simulation. Vertical line shows void radius decided by void finder. red: $30<R<35$ [Mpc]. green: %$35<R<40$ [Mpc]. blue: $40<R<45$ [Mpc]. Error bar shows 1$\sigma$.}
%\label{stackcon}
%\end{center}
%\end{figure}
%\afterpage{\clearpage}
 
%fitting with convergence map
\subsection{Fitting the convergence profile}
\label{fittingwithconvergencemap}
In order to compute convergence and tangential shear profiles of voids
analytically, we assume the double top-hat model presented in
sec.\ref{sec:voidmodel}.  We consider the source at $\mathrm{z_s}=1.0$
and the lens at $\mathrm{z_l}=0.5$. For our model, we 
%TH%set
treat $\delta$,
$\theta_1$ and $\theta_2$ in eq.\ref{convergence3} and
eq.\ref{tangentialshear3} as free parameters to fit stacked lensing
signals in our simulation. At first, we ignore shape noise and only
consider noises from LSS and structures of voids in the process of
fitting. Fitting is performed in the parameter range
$-1\leq\delta\leq0$, $0'\leq\theta_1\leq100'$,
$0'\leq\theta_2\leq300'$. We sum up $\chi^2$ over the radius range
$0'\leq\theta\leq90'$  for $15\leq R\leq30$ Mpc, and
$0'\leq\theta\leq180'$ for $30\leq R\leq45$ Mpc. 
 
\begin{figure*}
\subfigure{\includegraphics[width=0.92\columnwidth]{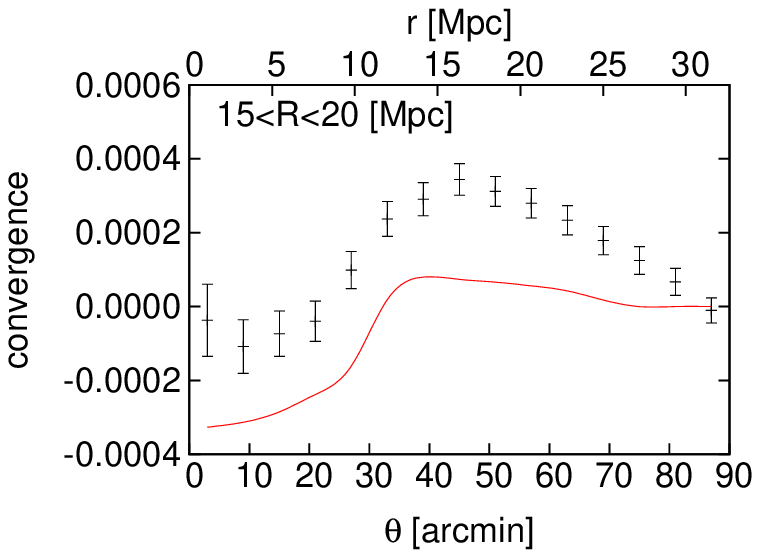}\label{con6-agam-1}}\hspace{1cm}
\subfigure{\includegraphics[width=0.92\columnwidth]{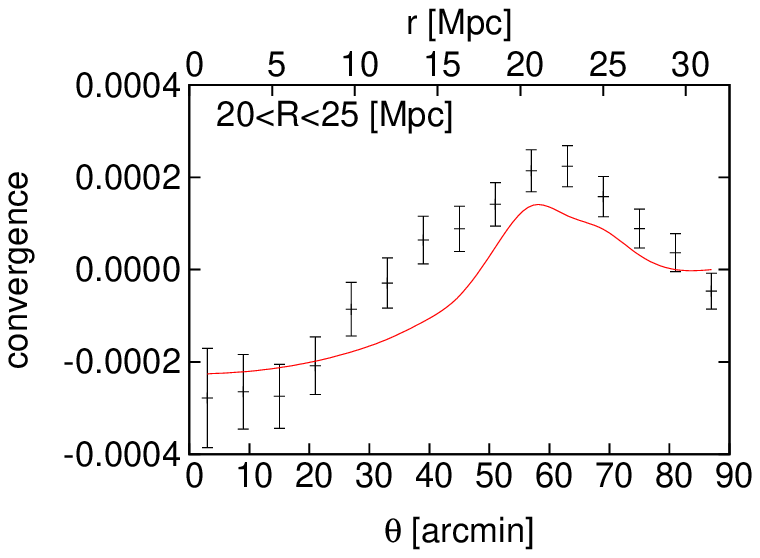}\label{con6-agam-2}}\hspace{1cm}
\subfigure{\includegraphics[width=0.92\columnwidth]{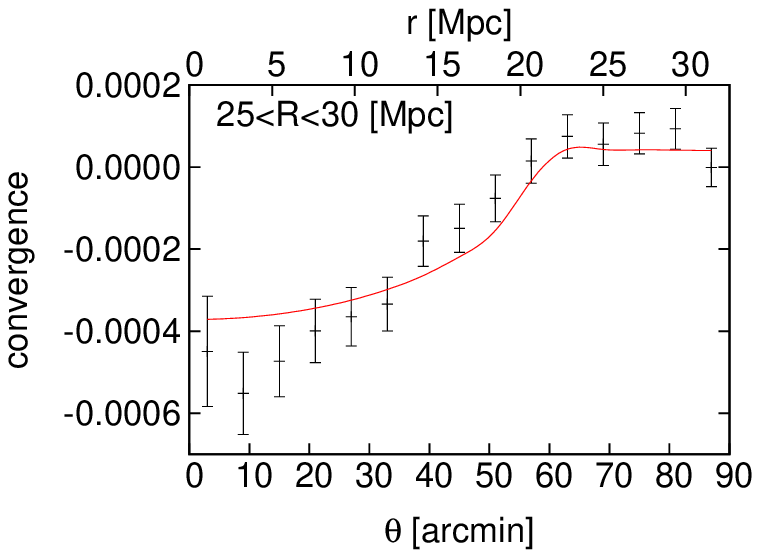}\label{con6-agam-3}}\hspace{1cm}
\subfigure{\includegraphics[width=0.92\columnwidth]{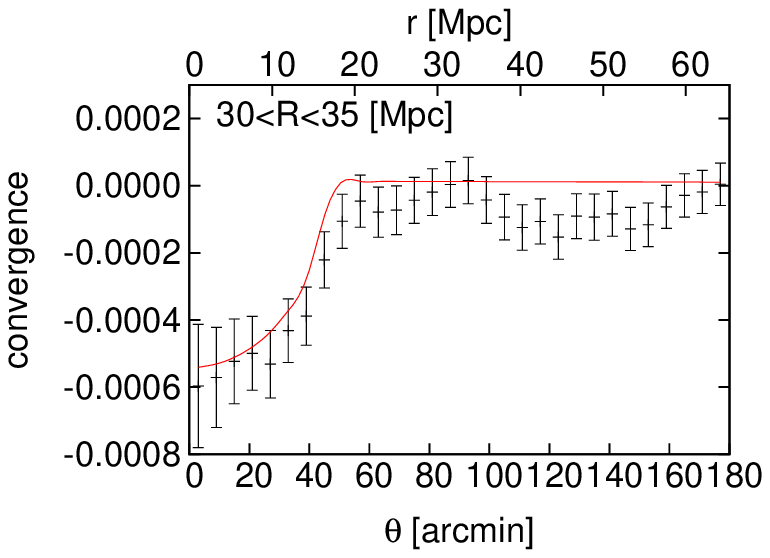}\label{con6-agam-4}}\hspace{1cm}
\subfigure{\includegraphics[width=0.92\columnwidth]{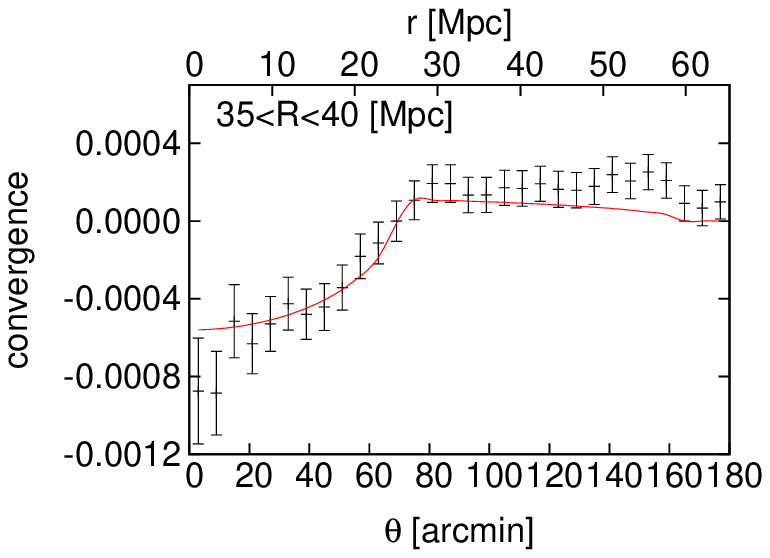}\label{con6-agam-5}}\hspace{1cm}
\subfigure{\includegraphics[width=0.92\columnwidth]{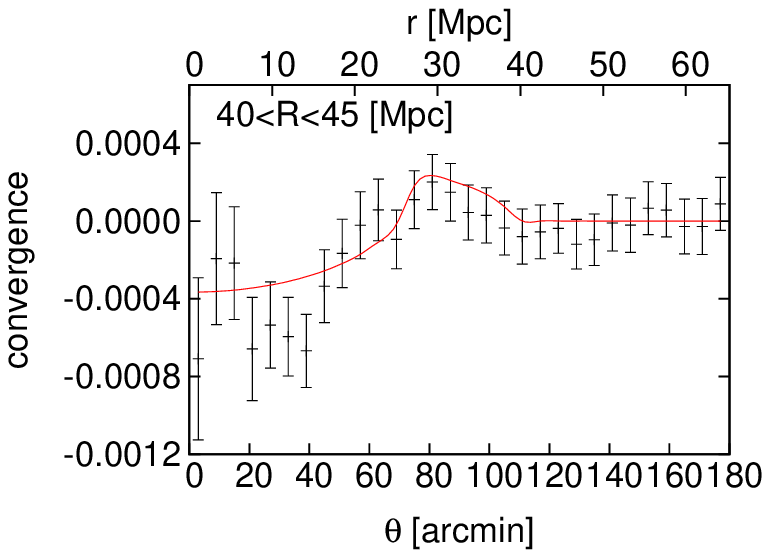}\label{con6-agam-6}}
\caption{Stacked convergence profiles estimated from simulation (points with errors) , the best-fit double top-hat model (solid line), assuming the survey area of $5000$ deg$^2$, the bin size $\Delta\theta=6$ arcmin and lens redshift $\mathrm{z_l}=0.5$. Stacking analysis are carried out in each radius derived from the void finder; $15\leq \mathrm{R} \leq20$ Mpc ({\it top-left}); $20\leq \mathrm{R} \leq25$ Mpc ({\it top-right}); $25\leq \mathrm{R} \leq30$ Mpc ({\it middle-left}); $30\leq \mathrm{R} \leq35$ Mpc ({\it middle-right}); $35\leq \mathrm{R} \leq40$ Mpc ({\it bottom-left}); $40\leq \mathrm{R} \leq45$ Mpc ({\it bottom-right}). Best-fit parameters are summarized in table.\ref{bestfitpara2}.}
\label{convergenceprofile1}
\end{figure*}

 Fig.\ref{convergenceprofile1} 
%TH%shows 
compares the convergence profiles from the simulation
 with the model. Model fitting is carried out for convergence
 profiles. We find that our simple model 
%TH%reconstructs 
reproduces lensing signals in
 simulation well. Best-fit parameters used in this plot are summarized in
 table.\ref{bestfitpara2}. Using these parameters, we estimate total
 void masses, i.e. the total mass removed from the void region
 (table.\ref{bestfitpara2}). Our model fits both convergence and
 tangential shear profile well, except for the case with $15\leq
 \mathrm{R}\leq20$ Mpc. There are two reasons why fitting fails for this
 case. The first is a problem in the Void Finder. 
As we and \citet{2009ApJ...699.1252F} showed, the number counts of
 voids depend on the minimum radius parameter $\xi$ of Void Finder
 especially for smaller radii, indicating that the Void Finder 
%fails to find true small voids
 mis-identifies spurious voids. 
The second reason is that the double top-hat model with the requirement
of the mass compensation may not be very realistic.
 
 Total void masses calculated by fitting the convergence
 profile with our model are a few times larger than masses estimated
 from direct integration at $\kappa(\theta)<0$. This is
 because observed convergence value is cumulative value from sources to
 an observer. Therefore when we calculate mass we should consider the
 effect from the ridge outside the void. Otherwise we underestimate
 total void mass. Also, the size of underdense regions $D_l\theta_1$ determined by
 model fitting are found to be smaller than radii defined in
 the Void Finder. This is probably because the halo catalogue we use for
 void finding does not exactly trace the underlying matter distribution.   
\begin{table*}
\begin{center}
\caption{Best fit parameters and derived total void mass. These parameters are determined from convergence profiles. Column (1): radius of void determined with the void finder; Column (2): density contrast; Column (3): radius of underdense region; Column (4): radius of overdense region; Column (5): void mass derived from the double top-hat model}
\label{bestfitpara2}
\begin{tabular}{cccccc}%{|c|c|c|c|c|}
\hline
Radius in the void finder [Mpc]&$\delta$&$\mathrm{D_l}\theta_1$ [Mpc]&$\mathrm{D_l}\theta_2$  [Mpc]&Void mass [$\mathrm{M}_{\odot}$]\\ \hline\hline
$15\sim20$ &-0.390&12.1&25.5&$1.46\times10^{15}$\\ %\hline
$20\sim25$ &-0.194&19.1&27.5&$2.83\times10^{15}$\\ %\hline
$25\sim30$ &-0.244&21.0&63.0&$4.78\times10^{15}$\\ %\hline
$30\sim35$ &-0.424&16.6&109&$4.09\times10^{15}$\\ %\hline
$35\sim40$ &-0.316&25.5&59.8&$1.10\times10^{16}$\\ %\hline
$40\sim45$ &-0.220&27.3&38.9&$2.73\times10^{16}$\\ 
\hline
\end{tabular}
\end{center}
\end{table*}

%\afterpage{\clearpage}
%\newpage

%fitting with tangential shear map
\subsection{Fitting the tangential shear profile}
\label{fittingwithtangentialshearmap}
 We also use tangential shear profile for finding best-fit parameters. We set same condition for searching best-fit parameters as that used for fitting the convergence profiles.
   
\begin{figure*}
\subfigure{\includegraphics[width=0.92\columnwidth]{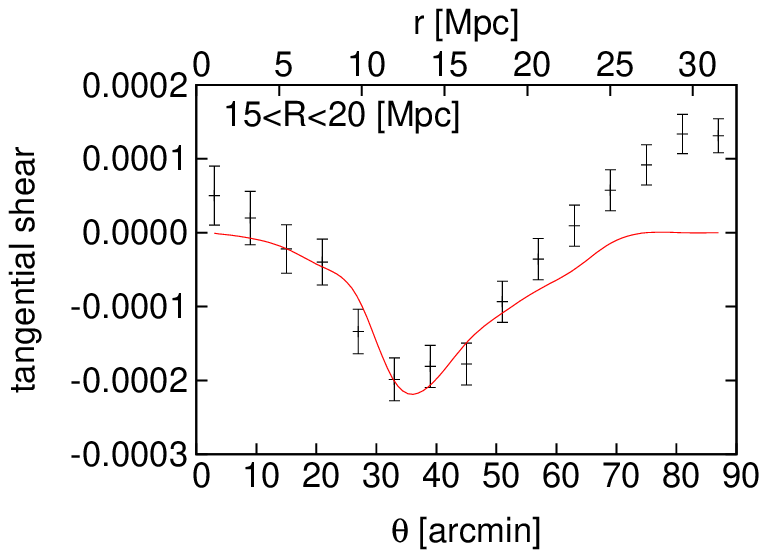}\label{agam6-con-1}}\hspace{1cm}
\subfigure{\includegraphics[width=0.92\columnwidth]{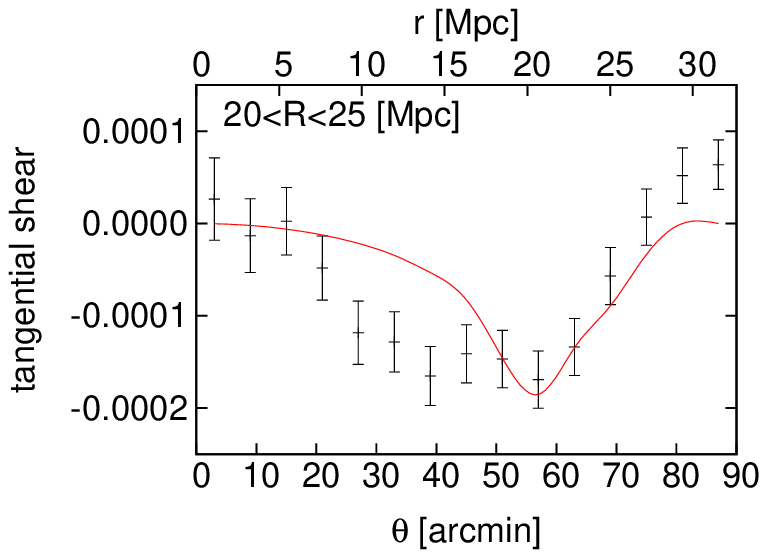}\label{agam6-con-2}}\hspace{1cm}
\subfigure{\includegraphics[width=0.92\columnwidth]{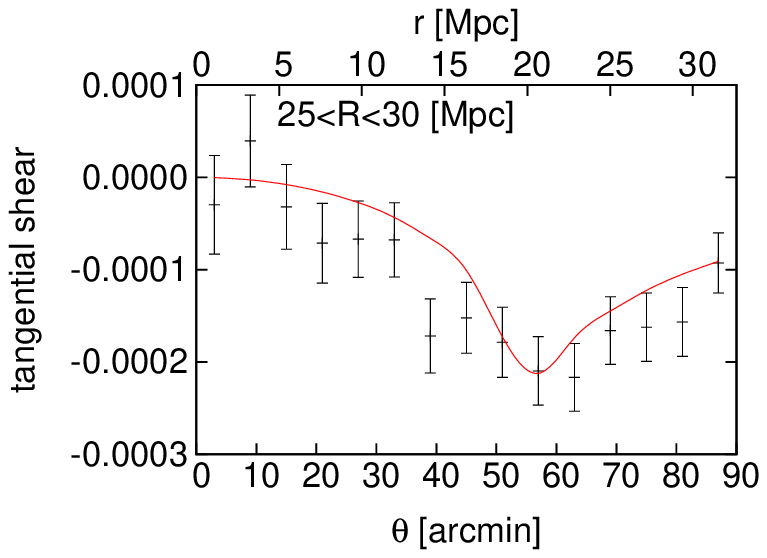}\label{agam6-con-3}}\hspace{1cm}
\subfigure{\includegraphics[width=0.92\columnwidth]{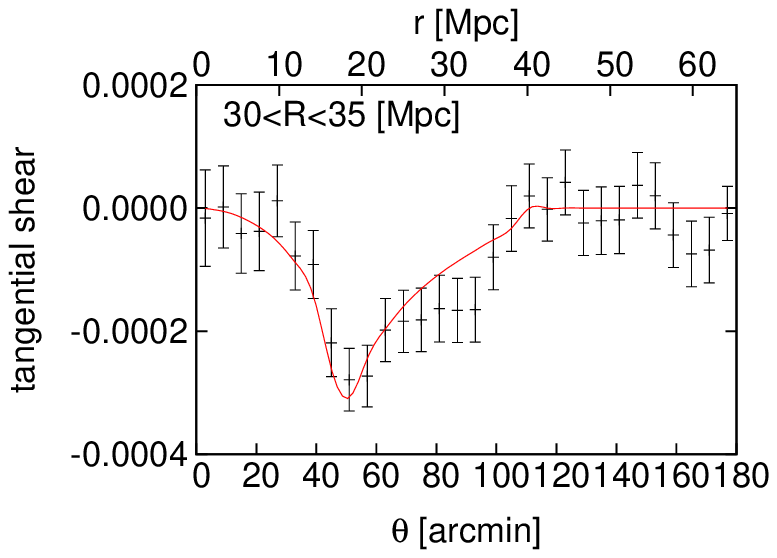}\label{agam6-con-4}}\hspace{1cm}
\subfigure{\includegraphics[width=0.92\columnwidth]{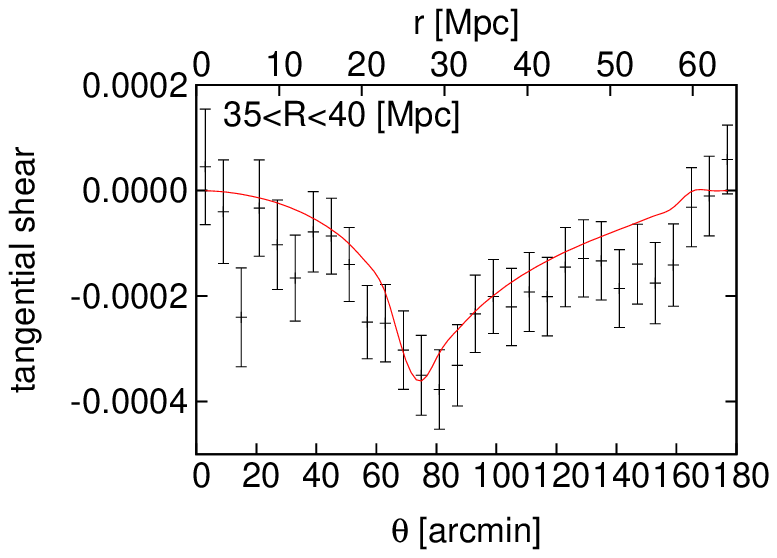}\label{agam6-con-5}}\hspace{1cm}
\subfigure{\includegraphics[width=0.92\columnwidth]{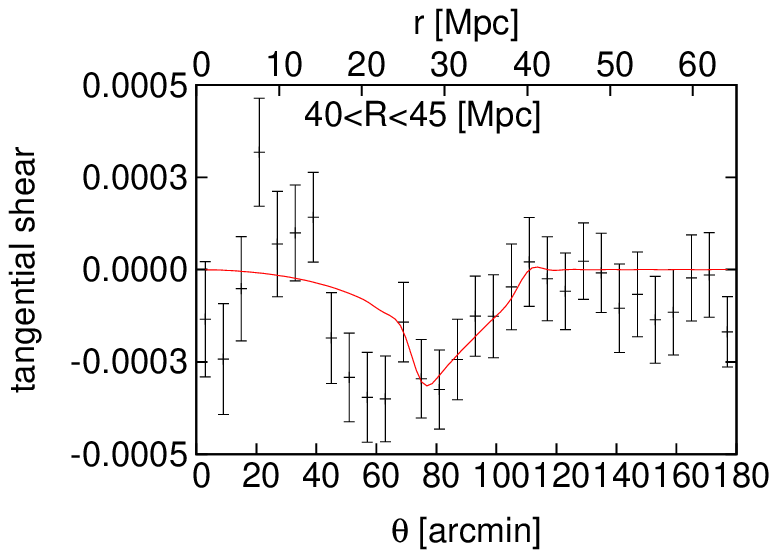}\label{agam6-con-6}}
\caption{Tangential shear profiles estimated from simulation (points with errors), the best-fit double top-hat model (solid lines) in each radius. Same assumptions in fig.\ref{convergenceprofile1} are adopted. Stacking analysis are carried out in each radius derived from the void finder; $15\leq \mathrm{R} \leq20$ Mpc ({\it top-left}); $20\leq \mathrm{R} \leq25$ Mpc ({\it top-right}); $25\leq \mathrm{R} \leq30$ Mpc ({\it middle-left}); $30\leq \mathrm{R} \leq35$ Mpc ({\it middle-right}); $35\leq \mathrm{R} \leq40$ Mpc ({\it bottom-left}); $40\leq \mathrm{R} \leq45$ Mpc ({\it bottom-right}). Best-fit parameters are summarized in table.\ref{bestfitpara4}.}
\label{tangentialshearprofile2}
\end{figure*}

 Figure.\ref{tangentialshearprofile2} summarizes tangential shear
 profiles in the simulation and best-fit
 models. Table.\ref{bestfitpara4} shows best-fit parameters and total
 void masses determined from the fitting. Total void masses are
 estimated with eq.\ref{massreconstruct2} as we have done in
 sec.\ref{fittingwithconvergencemap}. We find that the best-fit
 parameters and the resulting total void masses are very similar to
 those obtained from fitting convergence profiles
 (table.\ref{bestfitpara2}), suggesting the consistency of our
 analysis.%Using same parameters, we reconstruct convergence profiles
          %(fig.\ref{convergenceprofile2}).  

\begin{table*}
\caption{Best-fit parameters and derived total void mass. These parameters are determined from tangential shear profiles. Column (1): radius of void determined with the void finder; Column (2): density contrast; Column (3): radius of underdense region; Column (4): radius of overdense region; Column (5): void mass derived from the double top-hat model }
\begin{center}
\begin{tabular}{ccccc}
\hline
Radius in the void finder [Mpc]&$\delta$&$\mathrm{D_l}\theta_1$ [Mpc]&$\mathrm{D_l}\theta_2$  [Mpc]&Void mass [$\mathrm{M}_{\odot}$] \\ \hline\hline
$15\sim20$ &-0.398&12.2&25.3&$1.53\times10^{15}$\\ %\hline
$20\sim25$ &-0.170&19.1&27.6&$2.50\times10^{15}$\\ %\hline
$25\sim30$ &-0.250&19.0&109&$3.63\times10^{15}$\\  %\hline
$30\sim35$ &-0.403&16.8&40.0&$4.00\times10^{15}$\\ %\hline
$35\sim40$ &-0.290&25.5&59.8&$1.01\times10^{16}$\\ %\hline
$40\sim45$ &-0.183&27.3&39.5&$7.80\times10^{15}$\\ 
\hline
\end{tabular}
\end{center}
\label{bestfitpara4}
\end{table*}

%\afterpage{\clearpage}

\subsection{Signal-to-Noise ratio}
\label{signaltonoiselatio}
Using the stacked lensing covariance matrix we derived, we estimate S/N
%TH%for convergence and tangential shear. For tangential shear, S/N is
%estimated in the case of  with and without the shape noise. 
defined in eq.\ref{totalsn} for cases with and without the shape noise. 

\begin{figure}
\begin{center}
\includegraphics{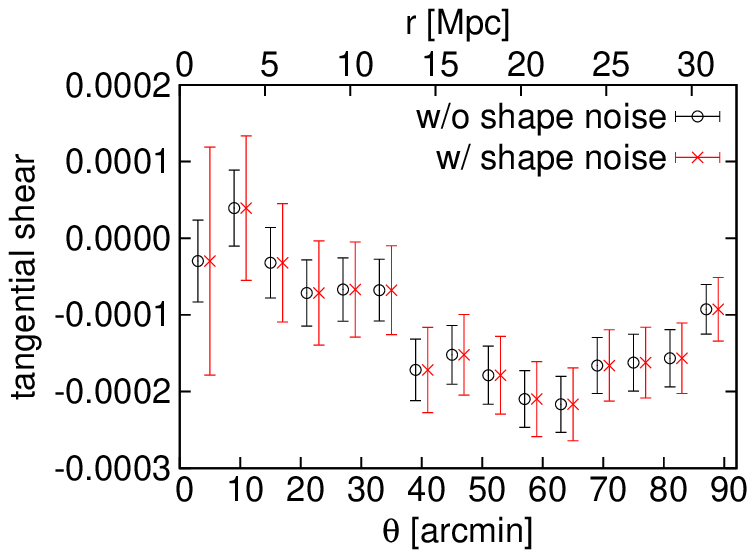}
\caption{Crosses and circles with errors show the tangential shear profiles for the cases  which include the shape noise and not. The staking analysis are carrie out for voids in the radial range $20\leq \mathrm{R}\leq25$ Mpc. We assume a HSC-type survey, $\mathrm{n_g}=30$ arcmin$^2$, $\sigma_\epsilon=0.4$ and FOV=$5000$ degree$^2$.}
\label{compareshapenoise}
\end{center}
\end{figure}

Fig.\ref{compareshapenoise} shows tangential shear profiles 
%TH%in both cases when the shape noise is and is not added to the
%covariance matrix. 
with error bars with and without the shape noise considered.
For illustrative purpose, points are shifted by $2$
arcmin. Stacking analysis are carried out for voids in the radial range
$20\leq \mathrm{R}\leq25$ Mpc and the redshift range
$0.4\leq\mathrm{z_l}\leq0.6$. The shape noise is estimated from
eq.\ref{shapenoise}. The error from the shape noise is added to the
diagonal elements of the covariance matrix. The effect from the shape
noise becomes a dominant component of the error at small scale, because
the number of galaxies in each radial annulus is proportional to the
radius under the condition that radial bin size $\Delta \mathrm{r}$ is
constant. 
%TH%Over the scale we consider, however, fig.\ref{compareshapenoise}
%shows that 
On larger scales, the shape noise is not a dominant component of the error. 

Total S/N is estimated for voids in the redshift ranges
$0.4 \leq \mathrm{z_l} \leq 0.6$ and $0.1 \leq \mathrm{z_l}\leq 0.3$ by using
eq.\ref{totalsn}. Table.\ref{sn1} shows S/N with and
without the shape noise for each radius. 
Table.\ref{sn2} shows S/N for the redshift range
$0.1\leq\mathrm{z_l}\leq0.3$. Table.\ref{sn1} and
table.\ref{sn2} show that the staked lensing signals are detected at
significant level of S/N$\geq5$. Adding the shape noise degrades S/N, but
the effect is relatively minor. Compared with total S/N in the redshift
range $0.4\leq\mathrm{z_l}\leq0.6$, total S/N in the redshift range
$0.1\leq\mathrm{z_l}\leq0.3$ are degraded. The error coming from the diversity
in the structure of each void increases because the number of voids
decreases as the redshift range becomes low. However, S/N for individual
void in the 
redshift range $0.1\leq\mathrm{z_l}\leq0.3$ is higher than that in the
redshift range $0.4\leq\mathrm{z_l}\leq0.6$ because of the larger apparent
sizes of voids at lower redshift. 

\begin{table*}
\caption{Signal-to-noise ratio, integrated over angular scales considered in the redshift range $\mathrm{z=0.4}\sim0.6$. Column (1): radius of void determined with the void finder; Column (2): the number count of voids in the each radial bin and the redshift range $\mathrm{z=0.4}\sim0.6$; Column (3): S/N derived from tangential shear without the shape noise; Column (4): S/N derived from tangential shear with the shape noise; Column (5): S/N for one void. It is estimated from information of tangential shear with shape noise and the number of voids (table.\ref{stackref1})} 
\begin{center}
\begin{tabular}{cccccc}
\hline
Radius [Mpc]&Number of voids&S/N (tangential shear&S/N (tangential shear&(S/N)/$\sqrt{N_{void}}$\\ 
&&w/o shape noise)&w/ shape noise)&(w/ shape noise)\\ \hline\hline
$15\sim20$&5246 &11.0&9.18&0.127\\ %\hline
$20\sim25$&3892 &8.25&6.97&0.112\\ %\hline
$25\sim30$&2446 &7.55&6.51&0.132\\  %\hline
$30\sim35$&1400 &8.19&6.76&0.181\\ %\hline
$35\sim40$&724 &7.57&6.25&0.232\\ %\hline
$40\sim45$&320 &7.89&5.39&0.301\\ 
%Radius [Mpc]&Number of voids&S/N&S/N (tangential shear&S/N (tangential shear&(S/N)/$\sqrt{N_{void}}$\\ 
%&&(convergence)&w/o shape noise)&w/ shape noise)&(w/ shape noise)\\ \hline\hline
%$15\sim20$&5246 &11.7&11.0&9.18&0.127\\ %\hline
%$20\sim25$&3892 &9.60&8.25&6.97&0.112\\ %\hline
%$25\sim30$&2446 &8.65&7.55&6.51&0.132\\  %\hline
%$30\sim35$&1400 &8.17&8.19&6.76&0.181\\ %\hline
%$35\sim40$&724 &8.08&7.57&6.25&0.232\\ %\hline
%$40\sim45$&320 &7.63&7.89&5.39&0.301\\ 
\hline
\end{tabular}
\end{center}
\label{sn1}
\end{table*}

\begin{table*}
\caption{Signal-to-noise ratio, integrated over angular scales
  considered in the redshift range $\mathrm{z=0.1}\sim0.3$. Column (1):
  radius of void determined with the void finder; Column (2): the number
  count of voids in the each radial bin and the redshift range
  $\mathrm{z=0.1}\sim0.3$; Column (3): S/N derived from tangential shear without the shape noise;
  Column (4): S/N derived from tangential shear with the shape noise;
  Column (5): S/N for one void}
\begin{center}
\begin{tabular}{cccccc}
\hline
Radius [Mpc]&Number of voids&S/N (tangential shear&S/N (tangential shear&(S/N)/$\sqrt{N_{void}}$\\ 
&&w/o shape noise)&w/ shape noise)&(w/ shape noise)\\ \hline\hline
$15\sim20$ &1798&6.89&6.09&0.144\\ %\hline
$20\sim25$ &1097&5.69&5.22&0.158\\ %\hline
$25\sim30$ &523&7.21&6.70&0.293\\  %\hline
$30\sim35$ &252&5.09&4.43&0.279\\ %\hline
$35\sim40$ &88&6.62&5.67&0.605\\ %\hline
%Radius [Mpc]&Number of voids&S/N&S/N (tangential shear&S/N (tangential shear&(S/N)/$\sqrt{N_{void}}$\\ 
%&&(convergence)&w/o shape noise)&w/ shape noise)&(w/ shape noise)\\ \hline\hline
%$15\sim20$ &1798&6.49&6.89&6.09&0.144\\ %\hline
%$20\sim25$ &1097&5.73&5.69&5.22&0.158\\ %\hline
%$25\sim30$ &523&7.36&7.21&6.70&0.293\\  %\hline
%$30\sim35$ &252&5.65&5.09&4.43&0.279\\ %\hline
%$35\sim40$ &88&6.93&6.62&5.67&0.605\\ %\hline
\hline
\end{tabular}
\end{center}
\label{sn2}
\end{table*}

%pressschechter theory
\subsection{Comparison of the mass function of voids}
\label{analyticalpredictionofnumbercountofvoids}
%TH% In order to predict the mass function of voids analytically, we
% determine the linear density fluctuation $\delta_\mathrm{v}$ from
% comparison of mass functions between the simulation and the modified PS
% theory (sec.\ref{pressschechter}).  In
% sec.\ref{fittingwithconvergencemap} and
% \ref{fittingwithtangentialshearmap}, masses of voids are estimated for
% each radius. From this and sec.\ref{sec:numbercount} the number counts
% of voids are obtained as a function of mass. We set $\delta_\mathrm{v}$
% as a free parameter. Fitting is performed in the parameter range
% $0.1\leq|\delta_\mathrm{v}|\leq1.0$.  
% 
% In order to obtain the mass function in each radius from the
% simulation, we use masses estimated from the tangential shear profiles
% (sec.\ref{fittingwithtangentialshearmap}). Because the inaccuracies of
% mass estimation from weak lensing signals affect the mass function, we
% fit masses as a function of the void radius assuming the following
% functional form

\begin{figure}
\begin{center}
\includegraphics{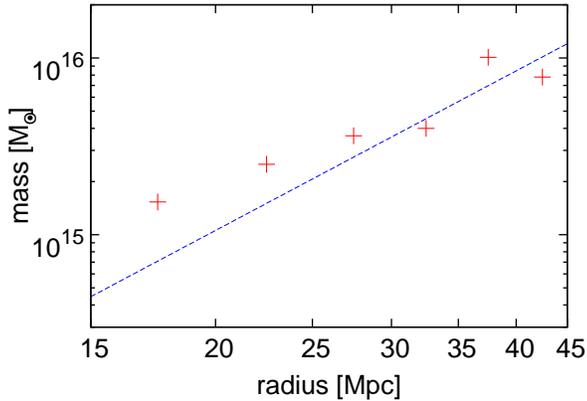}
\caption{Mass-radius relation. x-axis is radius obtained from void finder. y-axis is mass estimated from our model with tangential shear profiles. Crosses show masses obtained from weak lensing signals and the dashed line plots best-fit curve with eq.\ref{massrelation1}. }
\label{massrelation}
\end{center}
\end{figure}

\begin{figure}
\begin{center}
\includegraphics{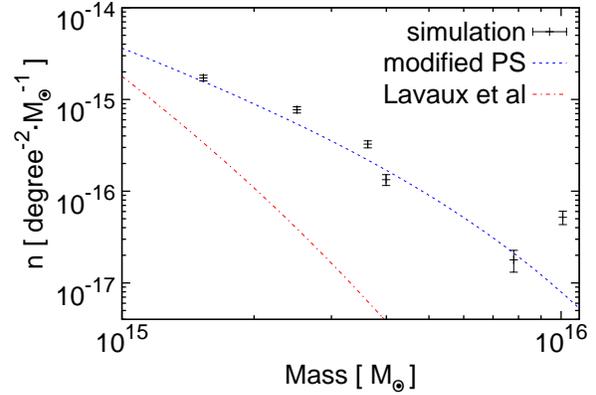}
\caption{Mass function of voids in simulation and the
    modified PS theory. Points with errors show the mass function of
    voids derived from the simulation. Masses are estimated from
    tangential shear profiles. The number count of voids in the redshift
    range $0.4\leq\mathrm{z}\leq0.6$ are used. Error bar shows 1$\sigma$
    of the number count with 200 realizations. The dashed line shows the
    best-fit curve in the modified PS theory with the linear density
    fluctuation of $\delta_v=-0.35$. For comparison, we also show the
    mass function estimated in \citet{2012ApJ...754..109L}.} 
\label{numbercountgamt}
\end{center}
\end{figure}

Here we derive the void mass function and fit it with the modified PS
model (sec.\ref{pressschechter}).
We consider voids in the redshift range of $0.4\leq z\leq0.6$.
The number counts of voids as a function of the void
radius are derived in sec. 6.1.
Thus, we have the void ``radius'' function.
In order to transform it to the ``mass'' function (namely, transforming
$\mathrm{n}(R)$ to $\mathrm{n}(M)$), we take the following approach;
The void masses for each radius group are evaluated from the tangential
shear profiles with the double top-hat model
(sec.\ref{fittingwithtangentialshearmap}).  
We assume the following relationship between the void radius and mass;
\begin{equation}
\mathrm{M}=\mathrm{A}_0\mathrm{R}^3,
\label{massrelation1}
\end{equation}
where $\mathrm{A}_0$ is constant and we treat it as a free
parameter. Fig.\ref{massrelation} shows the relationship obtained from
the simulation results along with the best-fit model which is determined
by the least-square method with $\mathrm{A}_0$ being 
\begin{equation}
\mathrm{A}_0=1.32\times10^{11}.
\label{massrelation2}
\end{equation}
Using eq.\ref{massrelation1}, we transform the radius interval to the
mass interval, and finally get the void mass function presented in Fig
\ref{numbercountgamt}.

The best fit modified PS void function is obtained by fitting the
simulation result with the analytical function, eq.\ref{ps4} where
$\delta_\mathrm{v}$ is treated as a free parameter.
We found the best fit model with $\delta_\mathrm{v}=-0.35$. 
The mass function from the simulation and the best-fit
  curve of the modified PS model are shown in
  figure.\ref{numbercountgamt}.  For checking the consistency, we also
  conduct same procedure using the results from the convergence profile,
  finding the best-fit linear density of $\delta_\mathrm{v}=-0.5$. We
  note that the parameter $\delta_\mathrm{v}$ estimated from our model
  differs from the value predicted from the spherical collapse
  model $\delta_\mathrm{v}=-2.81$ \citep{2004MNRAS.350..517S}. There are several possible reasons
  for this, including asphericity of voids and different
  definition of voids in our study from the spherical collapse model. 
To explore the origin of the discrepancy is beyond the scope of this
paper, and we leave it as future work.

In addition, we also compare the mass function from our
  simulation with one by \citet{2012ApJ...754..109L} in
  figure.\ref{numbercountgamt}. Our mass function is higher than that
  estimated in \citet{2012ApJ...754..109L}. 
The reason of this discrepancy is unclear. A possible reason is the
difference of void finding algorithms adopted in two studies as statistical
  properties of voids depend strongly on the definition of the void. 
  The other possibility is  the error of mass estimation. The different method to estimate the void mass may account for
  discrepancy. We leave further investigation on this discrepancy for
  future work.
%TH% The difference highlights the importance of the proper
%  calibration of void masses with stacked weak lensing, in order to make
%  unbiased comparions between theory and observations.} 

%Figure.\ref{numbercountgamt} shows the mass function in the simulation and the best-fit curve of the mass function in the modified PS theory. Mass functions of voids in simulation are reproduced with $\delta_\mathrm{v}=-0.35$.  For checking the consistency, we also do same works for the case of convergence. The best-fit linear density becomes as $\delta_\mathrm{v}=-0.5$ in that case. These values are smaller than the value predicted from the spherical top-hat model \citep{2004MNRAS.350..517S}. From our result, however, the linear density parameter is sensitive to the mass estimation reconstructed from weak lensing signals. For better understanding, we should study structures of voids and ridges more carefully. 

%%%%%%%%%%%%%%%%%%%%%%%%%%%conclusion%%%%%%%%%%%%%%%%%%%%%%%%%%%%%%%%%%%
\section{conclusion}
\label{sec:conclusion}
%TH% In this paper, we have presented the observability of voids with stacked
%weak lensing in next generation surveys such as the HSC survey. 
In this paper, we have examined a feasibility of detecting voids with
stacked weak lensing.
We select voids with a void finder from halo catalogue made from a large
set of $N$-body simulations. We have stacked convergence and tangential
shear data from the full ray-tracing simulations to obtain their
averaged radial profile (black in fig.\ref{convergenceprofile1} and
fig.\ref{tangentialshearprofile2}). From our stacking analysis, we have
seen both structures of void and ridge outside the void.

To fit the stacked lensing profiles obtained from the simulation, we
have considered a simple void model called double top-hat model. 
Our model fits both profiles of convergence and tangential shear in the
simulation very well (fig.\ref{convergenceprofile1} and
fig.\ref{tangentialshearprofile2}). Estimated total void masses from
this model were $\mathrm{M}=10^{14}\sim10^{16}\mathrm{M}_\odot$, which
were a few times larger than masses derived from direct integration of
the convergence profile at $\kappa(\theta)<0$. We have confirmed that
the dense ridges outside voids affect profiles of weak lensing signals
and the mass estimation of voids. This suggests that we have to properly
take account of this effect for interpreting stacked weak lensing signals
around voids. The radius of the underdense region derived from model
fitting is smaller than the radius derived from
the void finder, presumably because the halo catalogue used for finding
voids does not directly trace true dark matter distributions. 

%TH% This is not our finding %%% At small radii,
%the Void Finder cannot find true voids if the minimum radius $\xi$ is
%set to small value. This result is consistent with the previous paper
%\citep{2009ApJ...699.1252F}. 
We also derive the void mass function from simulation results, and
compared with the analytical mass function by the modified PS model.
%In order to predict the mass function of
%voids analytically, we compare the mass function of voids in the
%simulation with the modified PS theory and 
We found that the modified PS model with the threshold linear density
(but we treat it as a free fittting parameter) can reproduce the
simulation result.
In addition, we have
estimated total S/N to find that stacked lensing signals from voids can
be detected at significant level (S/N$\geq5$) for the $5000$ degree$^2$
area, even if the error coming from the shape noise is added. The shape
noise error becomes a dominant component only at the small scale and
therefore does not significantly degrade the total S/N.

Our work has demonstrated the observability of voids with the stacked
weak lensing method armed with a galaxy (or any tracer of the dark
matter distribution) redshift cataloges.
%We have also suggested a simple model to interpret
%the lensing signals. 
In this work, the average number density of halos
is $200$ /degree$^{2}/\Delta z$ for $\Delta z=0.1$ in the redshift range
$0.4<\mathrm{z}<0.5$. In the
SDSS-$\mathrm{I\hspace{-.1em}I\hspace{-.1em}I}$ Baryon Oscillation
Spectroscopic Survey (BOSS) \citep{2012arXiv1208.0022D}, the average
number density of luminous red galaxies (LRG) is about 50
/degree$^{2}/\Delta z=$ for $\Delta z=0.1$ in the same redshift range. 
The minimum void size we can identify is mainly determined by the average
distance between galaxies. In the simulation, it is about $6.86$
Mpc. In BOSS, it is about $10.9$ Mpc, which is thus about a factor of two
larger than the average distance used in this work. Therefore, it may be
possible to find voids from the BOSS data which have large radii, say
$>20$ Mpc, and to apply the technique described in this paper using weak
lensing measurements from e.g., HSC survey. This methodology can also be
applied to next generation surveys such as Euclid
\citep{2011arXiv1110.3193L}, 
Dark Energy Survey
(DES) \citep{2005astro.ph.10346T} and Large Synoptic Survey Telescope
(LSST) \citep{2009arXiv0912.0201L}.   

%%%%%%response to referee's comments%%%%%%%%%%%%%%%%%%%%%%%%%%%%%%
 The results presented in this paper can be used to
   estimate the detectability of 
   stacked lensing signal in the HSC survey in combination with the BOSS
   data. In the discussion above, we assumed that the number density of
   galaxies is $\mathrm{n_g}=30$ arcmin$^{-2}$ and the source redshift
   is $\mathrm{z_s}=1$. For the actual HSC survey, however, we can use
   about $80\%$ galaxies that are located behind $\mathrm{z_l}>0.6$ the
   voids we consider, which decreases the number density to
   $\mathrm{n_g}=24$ arcmin$^{-2}$. On the other hand, we assume a
   conservative error on the shape of each source galaxy,
   $\sigma_\epsilon=0.4$, and as a result the shot noise considered here
   is close to realistic estimates of the shot noise for the HSC survey
   (see, e.g. \citealt{2011PhRvD..83b3008O}). In this case we can simply
   scale S/N by the survey area. The survey area of the HSC survey is
   about $1400$ deg$^2$, indicating that the total S/N is degraded by a
   factor of $\sqrt{1400/5000}\sim 0.53$. Thus we expect the S/N from
   voids in the HSC survey is S/N$\geq3$ for each 
 void group classified by the void radius. Therefore it is possible to
 detect lensing signals from voids in 
 the HSC survey, particularly if we combine results for several void
 radius bins.

When we were writing this paper, we came across a paper by
\citet{2012arXiv1210.2446K} which also studied stacked weak lensing
signals around voids. The main difference of their paper from our
analysis is that \citet{2012arXiv1210.2446K} assumed analytic mass
profiles to estimate S/N, while in this paper we present realistic
stacked lensing profiles and the error covariance with ray-tracing of
$N$-body simulations. In addition, \citet{2012arXiv1210.2446K}
considered voids with smaller radii of $\leq15h^{-1}$Mpc for which our
void finder does not work well. Nevertheless, by extrapolating both the
results we argue that they are broadly consistent with each other.
We also confirmed that the density profile assumed in
  \citet{2012arXiv1210.2446K}, which originated from the density profile
  used in \citet{2012ApJ...754..109L}, also fits the stacked lensing
  profiles in our simulation well and produces similar void mass
  estimates.

%\afterpage{\clearpage}
%\newpage
%%%%%%%%%%%%%%%%%%%%%%%%%%%%acknowledgment%%%%%%%%%%%%%%%%%%%%%%%%%%%%%%%
\section*{Acknowledgments}
We thank an anonymous referee for useful comments. We would like to thank Masanori Sato for providing the raytracing data and  Yosuke Utsumi, Tsz Yan Lam, Masahiro Takada for useful discussions and comments. We also thank Caroline Foster for making the code VoidFinder freely available. Numerical computations in the paper were in part carried out on the general-purpose PC farm at Center for Computational Astrophysics, CfCA, of National Astronomical Observatory of Japan. This work was supported in part by the FIRST programme eSubaru Measurements of Images and Redshifts (SuMIRe)f, World Premier International Research Center Initiative (WPI Initiative), MEXT, Japan, and Grant-in-Aid for Scientific Re- search from the JSPS (23740161), and in part by Grant-in-Aid for Scientific Research from the JSPS Promotion of Science (23540324, 23740161)

%%%%%%%%%%%%%%%%%%%%%%%%%%%%reference%%%%%%%%%%%%%%%%%%%%%%%%%%%%%%%%%%%
\bibliographystyle{mn2e}
\bibliography{mn-jour,voidstack}

\end{document}